\documentclass[twocolumn,prc,showpacs,preprintnumbers,superscriptaddress]{revtex4-2}

\usepackage[T1]{fontenc}
\usepackage{amsmath} 
\usepackage{amssymb}
\usepackage{amsfonts}
\usepackage{mathrsfs}
\usepackage{units}
\usepackage{multirow}
\usepackage{color}
\usepackage{braket}

\usepackage{dcolumn}
\usepackage{bm}
\usepackage{rotating} 

\usepackage{rotating,epsfig,dcolumn}
\usepackage{graphicx,bm}
\usepackage{bm,color}
\usepackage{graphicx}
\usepackage{epstopdf}
\usepackage{rotating}
\usepackage{float}
\usepackage{multirow}
\usepackage{lipsum}
\usepackage[utf8]{inputenc}

\newcommand{\disregard}[1]{}

\newcommand{\bL}{\begin{Large}}
\newcommand{\eL}{\end{Large}}
\newcommand{\be}{\begin{equation}}  
\newcommand{\ee}{\end{equation}}
\newcommand{\ba}{\begin{eqnarray*}}
\newcommand{\ea}{\end{eqnarray*}}

\newcommand{\gras}[1]{\boldsymbol{#1}}

\newcommand{\bn}{\begin{eqnarray}}
\newcommand{\en}{\end{eqnarray}}
\newcommand{\bc}{\begin{center}}
\newcommand{\ec}{\end{center}}
\newcommand{\bi}{\begin{itemize}}
\newcommand{\ei}{\end{itemize}}

\newcommand{\ti}[1]{#1\index{#1}}

\renewcommand{\ti}[1]{#1}

\usepackage{tikz}
\newcommand{\tikzcircle}[2][red,fill=red]{\tikz[baseline=-0.5ex]\draw[#1,radius=#2] (0,0) circle ;}%

\newsavebox{\tmpstrikebox}
\newlength{\tmpstrikelen}

\begin{document}

\title{Sensitivity study of mirror energy differences in positive parity bands of $T=\frac{3}{2}$ A=45 nuclei.}

\author{W. Satu{\l}a}
\affiliation{Institute of Theoretical Physics, Faculty of Physics, University of Warsaw, ul. Pasteura 5, PL-02-093, Warsaw, Poland}
\author{M. A. Bentley}
\affiliation{School of Physics, Engineering and Technology, University of York, Heslington, York. YO10 5DD, United Kingdom}
\author{A. Jalili}
\affiliation{Institute of Theoretical Physics, Faculty of Physics, University of Warsaw, ul. Pasteura 5, PL-02-093, Warsaw, Poland}
\author{S. Uthayakumaar}
\affiliation{School of Physics, Engineering and Technology, University of York, Heslington, York. YO10 5DD, United Kingdom}
\affiliation{Facility for Rare Isotope Beams, Michigan State University, East Lansing, Michigan 48824, USA}

\date{\today}

\begin{abstract}{Symmetry conserving density functional theory (DFT) based no-core-configuration-interaction framework (DFT-NCCI) is an 
excellent tool for precision calculation of diverse (pseudo-)observables related to isospin symmetry breaking from  
elusive isospin impurities trough isospin corrections to superallowed beta decays to mirror- and triplet-displacement energies
and mirror energy differences (MED) along rotational bands.  In our recent work [Phys. Rev. C {\bf 106}, 024327 (2022)]
we performed axial DFT-NCCI calculations and failed to reproduce a sign of MED in positive-parity 
($\pi =+$) bands of $^{45}$Sc/$^{45}$Cr $T=3/2$ mirror pair what casts a shadow on credibility of the model. 
In this work we aim to perform a thorough analysis 
of this case with the focus on sensitivity of our predictions with respect to: ({\it i\/}) low-energy constants (LECs) of our effective contact charge symmetry breaking (CSB) force and ({\it ii\/}) nuclear shape. We demonstrate, among the other, that inclusion of 
triaxial $\pi =+$ ground-state - which is actually the global $\pi =+$ minimum in our unconstrained
mean-field calculation - in the DFT-NCCI calculations instead of the axial one used before leads to MED which are consistent 
with experimental data concerning both their sign as well as  magnitude without any need for fine-tuning of the model's LECs. 
}
\end{abstract}

\maketitle

\section{Introduction}

The origin of isospin symmetry breaking (ISB) depends on resolution of the underlying theory. 
 At the fundamental level of QCD/QED, where the degrees of freedom or, equivalently, 
the primary building blocks are quarks and gluons, the ISB originates from different 
masses (strong component) and  charges (electromagnetic component) of constituent {\it up\/} and {\it down\/} quarks.  
This relatively simple and transparent picture complicates quite radically when quarks and gluons are replaced by point-like 
nucleons interacting via intermediate mesons. The effective field theories or high-precision meson-exchange potentials contain several 
sources of ISB which are intertwined. They include, to name a few, nucleon mass splitting, one- and two-boson exchange
terms (2$\pi$ exchange with the intermediate $\Delta$, $\pi\rho$, $\rho\omega$), pion mass splitting or $\pi\gamma$ exchange.
All these terms, following the work of Henley and Miller~\cite{(Hen79)}, can be grouped into 
three distinct classes: class II (isotensor or charge-independence-breaking (CIB)), class III (isovector or charge symmetry 
breaking (CSB)), and class IV forces. The parameters of high-precision potentials are 
fitted directly to phase shifts (and selected two- and three-body data) and are used subsequently to compute 
finite nuclei using advanced many-body techniques which do not break manifestly fundamental symmetries.
Such methodology is commonly known as {\it ab initio\/} approach to nuclear structure.

Further reduction of degrees of freedom to densities and currents generated by point-like nucleons brings us 
to the level of approximation commonly known as density functional theory (DFT).  The primary object of 
interest here is an energy density functional (EDF). In nuclear physics one often explores 
formal  similarity between the Hartree-Fock and Kohn-Sham schemes building the nuclear EDF by means 
of Hartree-Fock(-Bogolyubov) technology with effective phenomenological EDF generators 
like the explored hereafter Skyrme interaction which leads to local DFT. Such formalism, called single-reference
DFT (SR-DFT) is a method of choice to compute globally bulk nuclear observables like binding energies, radii,
quadrupole moments or rotational inertia parameters.

The inherent feature of nuclear SR-DFT is a mechanism of spontaneous symmetry breaking (SSB). 
It is in fact a key to its success in computing bulk observables but, simultaneously, 
it is a feature that hampers applicability of the method to, in particular, computation 
of energy spectra or transition rates. This deficiency can be cured by restoring broken symmetries 
with the use of projection techniques i.e. by generalizing the SR-DFT to the so called multi-reference 
DFT (MR-DFT). While the former operates with a single reference Slater determinant $|\varphi\rangle$, the latter uses 
a linear combination of Slater determinants rotated in space (isospace, gauge space) $\hat R |\varphi\rangle$ 
with weights determined by the symmetry group.  Furthermore, after mixing states  projected from 
different quasi-particle or particle-hole configurations one is able to reach with these techniques a level 
of functionality comparable to the nuclear shell-model (NSM). 

There are different realizations of DFT-based configuration-interaction methods, see Ref.~\cite{(She21)}. 
The Warsaw group has developed the DFT-NCCI (no-core-configuration-interaction) variant based on the unpaired Skyrme 
functional that includes CSB and CIB terms up to next-to-leading (NLO) order
and a unique combination of the angular-momentum and isospin projections with an aim to 
study ISB in $N\approx Z$ nuclei. With this method we were able to compute different 
observables and pseudo-observables including: isospin impurities~\cite{(Sat09)}, ISB corrections
to the superallowed $0^+ \rightarrow 0^+$~\cite{(Sat11)} and $T=1/2$ mirror~\cite{(Kon22)} 
beta decays as well as mirror- (MDE) and triplet-displacement energies (TDE)~\cite{(Bac18),(Bac19)}
with the accuracy comparable to the NSM, see Refs.~\cite{(Har15),(Har20),(Nav09a),(Gon19),(Orm89)} and references quoted therein. 
Recently, we have also applied the DFT-NCCI formalism 
to compute mirror energy differences (MED) which are defined as follows:
\begin{equation}
    MED_J = E^{*}_{J,T,-T} - E^{*}_{J,T,T}
\end{equation}
where $E^{*}_{J,T,T_z}$ is the excitation energy of a particular state that has a total angular momentum (spin) $J$, isospin $T$ and isospin projection $T_z$. 
From theoretical perspective MED are very demanding quantities as they are very sensitive to details of the changes in configurations in function of increasing excitation energy and angular momentum. In spite of that we were able to reach reasonable agreement
in $T=1/2$ mirrors from the lower $fp$-shell~\cite{(Bac21)}, the $T=3/2$ $^{47}$Ti/$^{47}$Mn mirrors~\cite{(Uth22)}, and
the  very heavy $T=1/2$ $^{79}$Zr/$^{79}$Y 
mirrors~\cite{(Lle20)} without adjusting locally a single coupling constant. 
We failed, however, to reproduce MED for unnatural ($\pi=+$) parity bands in $^{45}$Sc/$^{45}$Cr. 
The aim of present work is to investigate sensitivity of DFT-NCCI calculations for MED in $\pi=+$ parity bands in $^{45}$Sc/$^{45}$Cr 
with respect to low-energy couplings (LECs) of the local CSB force and with respect to shape degrees of freedom. 
We report here the first DFT-NCCI calculations for MED that admit not only axial but also traxial configurations in 
the model's configuration space.
The paper is organized as follows. In Sect.~\ref{sec:ncci} we briefly introduce the DFT-NCCI model. 
In Sect.~\ref{sec:LEC} we discuss sensitivity of MED in $\pi=+$ parity bands in $^{45}$Sc/$^{45}$Cr  with respect to 
LECs of the CSB force.  In Sect.~\ref{sec:tri} we present the results of calculations that include triaxial 
configurations. The paper is briefly summarized in Sec.~\ref{sec:sum}

\section{The DFT-NCCI approach}\label{sec:ncci}

The self-consistent Hartree-Fock(-Bogliubov) framework is a  powerful approach offering simple 
understanding of complex features of nuclear structure in terms of very intuitive deformed  independent-particle 
configurations.  It is therefore not surprising that it serves as a starting point for various more advanced theories 
which, in general, account for interactions between independent-particle configurations provided by mean-field. 
The  DFT-NCCI approaches, see  Ref.~\cite{(She21)} and refs. quoted therein for recent overview of different realizations 
of DFT-NCCI schemes, are good examples of such beyond-mean-field methods. These approaches can be 
characterized as post Hartree-Fock(-Bogoliubov) configuration-interaction methods that aim to restore 
symmetries violated by mean-field and mix good symmetry states projected from different mean-field configurations. 

We developed recently the DFT-NCCI variant dedicated to study isospin-symmetry-breaking 
phenomena in $N\sim Z$ nuclei.  The model is based upon unpaired charge-dependent local EDF 
generated by density-independent Skyrme pseudo-potential and involves a unique combination 
of the angular-momentum and isospin projections. In practical calculations we use 
SV$^{\rm ISB}_{\rm T;\, NLO}$ Skyrme pseudo-potential that includes density-independent isoscalar Skyrme pseudo-potential SV of Ref.~\cite{(Bei75)} [albeit with tensor terms included in the SV-EDF for the sake of mathematical consistency] 
augmented  with class~III CSB interaction:~\footnote{The class~II CIB force is inactive in isospin doublets discussed here
and will be therefore omitted.} 
\begin{eqnarray}
\hat{V}^{\rm{III}}(i,j)  =  \bigg[&&
t_0^{\rm{III}}  \delta\left(\gras{r}_{ij} \right)
  +  \frac12 t_1^{\rm{III}}
\left( \delta\left( \gras{r}_{ij} \right) \bm{k}^2 + \bm{k}'^2 \delta\left(\gras{r}_{ij} \right) \right)   
\nonumber \\
& &+  t_2^{\rm{III}}
\bm{k}' \delta\left(\gras{r}_{ij} \right) \bm{k} \bigg]  \left( \hat{\tau}_3^{(i)}+\hat{\tau}_3^{(j)} \right) ,
\label{eq:classIII_NLO}
\end{eqnarray}
where $\gras{r}_{ij} = \gras{r}_i - \gras{r}_j$,
$\bm{k}  =  \frac{1}{2i}\left(\bm{\nabla}_i-\bm{\nabla}_j\right)$ and
$\bm{k}' = -\frac{1}{2i}\left(\bm{\nabla}_i-\bm{\nabla}_j\right)$ are the standard relative-momentum
operators acting to the right and left, respectively.
The three new  LECs are  equal: $t_0^{\rm{III}}= {}$\mbox{$11\pm2$\,MeV\,fm$^3$},
   $t_1^{\rm{III}}= {}$\mbox{$-14\pm4$\,MeV\,fm$^5$}, and $t_2^{\rm{III}} = {}$\mbox{$-7.8\pm0.8$\,MeV\,fm$^5$}.
They have been adjusted globally to all available data on MDEs for $A\geq 6$ in Ref.~\cite{(Bac19)}, which makes   
our approach free from adjustable parameters. 


The model will be applied to compute MED. As mentioned above, our method allows for rigorous treatment 
of both rotational and isospin symmetries. We have verified, however, that for the case of positive-parity bands in $A=45$ 
mirrors (in the calculation based on axial configurations), the spurious isospin mixing~\cite{(Sat09a)} very weakly affects 
the calculated MED and can be safely omitted.  Hence, similar to all other applications of the DFT-NCCI to MED, we shall restrict
ourselves to angular-momentum projection only. However, at variance to all other applications of the DFT-NCCI to MED we shall 
admit, in the final calculations, also triaxial configurations.

For the sake of completeness let us recall the model's computational scheme. It proceeds in three major steps: 
\begin{enumerate}
\item
First, we compute the so called a {\it configuration space\/}. It consists a set of 
$N_{\rm conf}$ relevant low-lying (multi)particle-(multi)hole self-consistent Hartree-Fock 
solutions $\{ \ket{\varphi_j}\}_{j=1}^{N_{\rm conf}}$ which are used in subsequent projection.
\item
Second, we apply the  angular-momentum  projection to each configuration
$\{ \ket{\varphi_j}\}$ separately in order to determine the family of states 
$\hat{P}^{I}_{MK}\ket{\varphi_j}$ having good angular momentum $I$ and it's projection 
on the intrinsic axis $K$. Since $K$ is, in general, not conserved we perform also
$K$-mixing, which gives us a set of good-angular-momentum states 
$\ket{\varphi_j ; I M; T_z}^{(i)}$, which form the so called {\it model space\/}:
\begin{equation}\label{m-space}
\ket{\varphi_j ; I M; T_z}^{(i)}=\frac{1}{\sqrt{\mathcal{N}^{(i)}_{\varphi_j ; IM;T_z}}}
\sum_{\substack{K }} a_{K}^{(i)} \hat{P}^{I}_{MK}\ket{\varphi_j}
\end{equation} 
where $K$ stands for a projection of angular momentum onto the intrinsic $z$-axis while  
\begin{eqnarray}
\hat P^I_{M K} & = & \frac{2I+1}{8\pi^2 } \int d\Omega\;
\; D^{I\, *}_{M K}(\Omega )\; e^{-i\gamma \hat{J}_z}
e^{-i\beta \hat{J}_y} e^{-i\alpha \hat{J}_z} ,
\end{eqnarray}
is the standard angular-momentum projection operator.
\item
Finally, we perform the mixing of, in general, non-orthogonal states
$\{\ket{\varphi_j ; I M; T_z}^{(i)}\}$ for all configurations $\{ \ket{\varphi_j}\}$ by 
solving the Hill-Wheeler equation. In effect, we obtain a set of linearly independent 
DFT-NCCI eigenstates of the form:
\begin{equation}
\ket{\psi_{\textrm{NCCI}}^{k; IM; T_z}}
=\frac{1}{\sqrt{\mathcal{N}^{(k)}_{IM;T_z}}}
\sum_{ij}c_{ij}^{(k)}\ket{\varphi_j;I M;T_z}^{(i)} \,, \label{eq:nccistate}
\end{equation}  

\end{enumerate}

Triaxial configurations imply  non-conservation of the intrinsic quantum number $K$
which, in turn, affects accuracy and stability of the calculations.  The problem 
of \ti{$K$-mixing} is handled in our code by solving, for each spin $I$ and 
each configuration $|\varphi\rangle \in \{ |\varphi_j\rangle \}_{j=1}^{j=n}$ separately, the non-orthogonal 
Hill-Wheeler (H-W) eigenvalue problem:  
\be\label{egn:HW}
\sum_{K'} H^{I}_{K K'} g^{(i)}_{IK'} = E_I^i \sum_{K'} N^{I}_{K K'} g^{(i)}_{IK'},
\ee
where
\bn\label{eqn:ker}
H^{I}_{KK'} & = & \langle \varphi | \hat H \hat{P}^I_{KK'} | \varphi \rangle , \\
\label{eqn:ovr}
N_{KK'} & = & \langle \varphi | \hat{P}^I_{KK'} | \varphi \rangle ,
\en
denote the Hamiltonian and norm kernels, respectively. The {\it model space\/} is   
overcomplete. We handle this problem by solving the H-W equation 
(\ref{egn:HW}) in the {\it collective basis\/}, spanned
by the {\it natural states\/}:
\begin{equation}   \label{nat_st}   |\varphi_j; \, IM; T_z\rangle^{(m)} =
  \frac{1}{\sqrt{n_m}} \sum_{K} \eta_K^{(m)}
  |\varphi_j; \,IMK; T_z\rangle.
  \end{equation}
The {\it natural states\/} used to construct the {\it model space\/} are the eigenstates of the 
norm matrix:
\be\label{egn:eignorm}
\sum_{K'} N_{K K'} \bar{\eta}^{(m)}_{K'} = n_m\; \bar{\eta}^{(m)}_{K},
\ee
having eigenvalues $n_m > \zeta$ larger than a certain 
cut-off parameter $\zeta$. In the present calculation we set $\zeta = 0.01$,
which guarantees numerical stability of the method. More details concerning
\ti{$K$-mixing} can be found in Ref.~\cite{(Dob09d)}.

The states (\ref{m-space}) spanning the {\it model space\/} are non-orthogonal. The final results are therefore 
calculated by solving again the H-W equation. At this stage we use the same technique of handling overcomplete
bases as outlined above for the case of \ti{$K$-mixing} i.e. we compute the eigenvalues and eigenstates 
of the norm matrix and construct the {\it natural states\/} and, in turn, the  {\it collective basis\/}. In this case 
we fix the cut-off parameter to be $\chi = 0.01$ in $^{45}$Sc and readjust it's value in $^{45}$Cr in order to obtain 
the collective basis of the same size in both nuclei.

\section{DFT-NCCI results for MED in $^{45}$Sc/$^{45}$Cr - sensitivity study}\label{sec:sens}

The DFT-NCCI method used hereafter has been defined in Ref.~\cite{(Sat16d)}, see also the supplemental
material to Ref.~\cite{(Uth22)}. It is the configuration-interaction framework with configuration space 
which is not fixed like in the conventional NSM but built step-by-step by adding physically-relevant low-lying (multi)particle-(multi)hole  mean-field configurations until reaching acceptable stability for the calculated observables. 
Building the configuration space we define first the active Nilsson orbitals $|N n_z \Lambda\, \Omega \rangle$, 
which are relevant for a given problem, and explore low-lying configurations built upon these orbitals. 
The positive parity bands discussed here 
involve particle-hole excitation(s) across the $N=Z=20$. Hence,  the active orbitals in our case include particle-like orbitals  
$|330\frac{1}{2}; \pm i\rangle$, $|321\frac{3}{2}; \pm i\rangle$,  $|312\frac{5}{2}; \pm i\rangle$, and $|303\frac{7}{2}; \pm i\rangle$ originating from the spherical $0f_{\frac{7}{2}}$-shell and hole-like orbitals
$|211\frac{1}{2}; \pm i\rangle$ and $|202\frac{3}{2}; \pm i\rangle$ originating  from the spherical 
$0d_{\frac{3}{2}}$  and the $|200\frac{1}{2}; \pm i\rangle$ Nilsson orbital 
originating from the spherical $1s_{\frac{1}{2}}$-subshell. The additional quantum number $r=\pm i$ denotes signature.
It reflects the fact that signature symmetry (and parity)  were superimposed on our mean-field solutions

All calculations presented below were done using 
a developing version of the HFODD solver~\cite{(Dob09d),(Sch17),(Dob21)}.  In the 
calculations, we used spherical basis consisting 12 harmonic oscillator shells.  
The integration over the Euler angles $\Omega = (\alpha, \beta, \gamma )$  is performed using the 
Gauss-Chebyshev (over $\alpha$ and $\gamma$) and Gauss-Legendre (over $\beta$) quadratures  
with $n_\alpha=n_\beta=n_\gamma=40$ knots.

The principles of our DFT-NCCI approach were laid down in Ref.~\cite{(Sat16d)}. 
Quantitative applications of the method to compute MED  became possible 
after implementing into the formalism contact CSB terms and adjusting 
their LECs to MDEs, see Refs.~\cite{(Bac18),(Bac19)}.  
Indeed, it is well known mostly from the earlier NSM studies, see Refs.~\cite{(Col98),(Orm89a),(Bro00a)} that these non-coulombic 
sources of ISB are critical both to cure the so called Nolen-Schiffer anomaly~\cite{(Nol69)} of MDEs as well as 
in quantitative description of MED versus $J$~\cite{(Zuc02),(Ben07),(Kan13),(Kan14),(Ben15)}. 
The very first calculation of MED with the use of DFT-NCCI approach was communicated 
by Llewellyn \emph{et al.} \cite{(Lle20)} who applied the method to the heaviest mirror pair studied so far, the $T_{z}=\pm\frac{1}{2}$ $A=79$ mirrors. In Ref.~\cite{(Bac21)} we applied
the method to the $T_{z}=\pm\frac{1}{2}$ mirrors from the lower $fp$-shell nuclei. In this work
the DFT-NCCI results were bench-marked with the shell-model results of Ref.~\cite{(Ben15)}. 
Recently, we applied the same method to the $T_{z}=\pm\frac{3}{2}$ mirror pairs in 
$^{45}$Sc/$^{45}$Cr and $^{47}$Ti/$^{47}$Mn. In all these applications we limited 
ourselves to axial mean-field configurations reaching, in general, reasonable agreement
with experimental data with the exception of positive parity band in $^{45}$Sc/$^{45}$Cr
mirrors where the DFT-NCCI failed to reproduce the overall sign of the MED.

\subsection{Sensitivity to LECs of the contact CSB force}\label{sec:LEC}

\begin{figure}[ht!]
\centering
\includegraphics[scale=0.50]{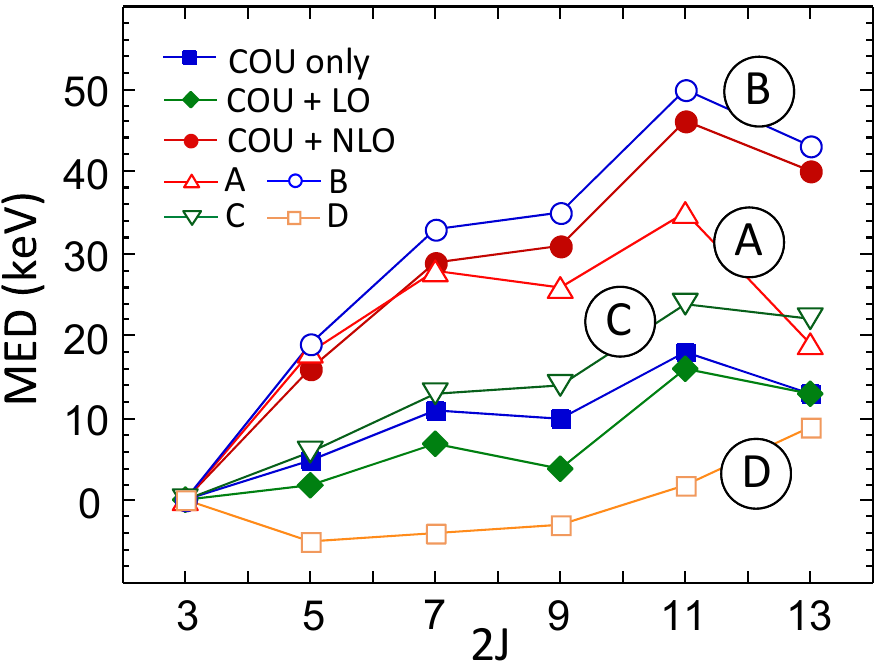}
\caption{(Color online) A test calculation of MED (see text) for the positive parity states in $^{45}$Sc/$^{45}$Cr in function of spin $2J$. Full symbols show calculations for three different variants of isospin-breaking force with optimal parameters of the 
local CSB interaction adjusted to MDEs. The variants include: the Coulomb force (blue squares), the Coulomb plus nuclear LO CSB (green diamonds) of Ref.~\cite{(Bac18)}, and  the Coulomb plus nuclear NLO CSB (red dots) of Ref.~\cite{(Bac19)}. 
Open symbols shows the results of tests performed using A, B, C, and D variants of the NLO CSB force, see the legend and 
Tab.~\ref{tab:new-nlo} for details concerning parameters of the NLO CSB forces used in the test. In the test calculation we use the same 11 low-lying configurations as in 
the analogous figure published in Ref.~\cite{(Uth22)}. } 
\label{fig:CSB-test}
\end{figure}

In Ref.~\cite{(Uth22)} we performed test calculation for MED in positive parity band 
including 11 most important  axial configurations and three different variants of the 
model including: 1) the Coulomb 
interaction alone; 2) the Coulomb plus LO CSB contact force~\cite{(Bac18)})  
the Coulomb plus NLO CSB force~\cite{(Bac19)}. These are reproduced here in Fig.~\ref{fig:CSB-test} (filled symbols). An unexpectedly strong 
effect of the NLO CSB force on the calculated MED is apparent, capable, in principle, of overturning 
its sign.  This prompted us to formulate in Ref.~\cite{(Uth22)} a conjecture that the data 
on MED may offer new opportunities to fine tune low-energy coupling 
constants of the NLO CSB force. Indeed, a fit to MDEs establishes only an overall magnitude 
of the CSB force without giving an access to its matrix elements which enter 
the theory through configuration-mixing. Moreover, even the static fit to MDEs
is not unique in a sense that the resulting penalty function exhibits quite 
pronounced softness along the line correlating two out of three LECs~\cite{(Bac19)}: 
\begin{equation}\label{soft}
t_1^{\rm{III}} \approx a(t_0^{\rm{III}} - b) - c
\end{equation}
where $a$=$-$1.65\,fm$^2$, $b$= 11\,MeV\,fm$^3$ and $c$=14\,MeV\,fm$^5$.  
In order to exploit the conjecture formulated in Ref.~\cite{(Uth22)} we 
decided to create six variants of the NLO CSB interaction and perform a sensitivity 
study for MED. The LECs of these new forces are collected in Tab.~\ref{tab:new-nlo}. The variants 
A and B follow the trend given by Eq.~(\ref{soft}) and have the original value of 
$t_2^{\rm{III}}$. Note that, in these two cases, the mirror  displacement energies (MDE) are very close to the
experimental value of 24\,MeV. In the variants C and D we fix original 
values for $t_0^{\rm{III}}$ and $t_1^{\rm{III}}$ and vary $t_2^{\rm{III}}$. 
In these cases we completely (albeit intentionally) deteriorate the agreement between theoretical and 
experimental MDEs. The variants E and F have all LECs different than original values. The calculated 
MDEs are again much worse than for the original force.

\begin{table}[t]
\centering
\caption{
Parameters of the NLO CSB forces used to perform sensitivity study of MED with respect to LECs of the effective
contact CSB force. Last column shows the calculated values of mirror displacement energy (MDE) defined
as binding-energy difference between  the $J=3/2^+$ band-heads of positive parity bands. 
Its experimental value is 24\, MeV.
 }
\label{tab:new-nlo}
\begin{tabular}{c|c|c|c|c}
 NLO       &  $t_0^{\rm{III}}$ & $t_1^{\rm{III}}$ & $t_2^{\rm{III}}$ & MDE  \\
Variant    &   [MeV\, fm$^3$]   &  [MeV\, fm$^5$]   & [MeV\, fm$^5$]  &  [MeV]  \\
\hline
A            &       -6.0                &        14.0              &      -7.8               &    24.126   \\
B            &        1.0                &          2.5              &      -7.8               &    24.104  \\
C            &       11.0               &         -14.0            &        0.0              &    21.705  \\
D            &       11.0               &         -14.0            &       7.8               &    19.333  \\
E            &       -6.0                &         -7.0              &       0.0               &    24.667  \\
F            &       -6.0                &         -7.0              &       3.0               &    23.753  \\
\end{tabular}
\end{table}


The MED calculated using test forces A, B, C, and D are shown in Fig.~\ref{fig:CSB-test} (open symbols). The results for the variants 
E and F appears to be quantitatively similar to the results obtained with the forces C and D, respectively.
Hence we refrain from showing them for the sake of clarity. The calculations clearly reveal strong dependence of MED 
on the gradient-dependent terms. It is rather clear, however, that the NLO force is not capable to change the sign, 
definitely not without completely deteriorating MDE.

\subsection{Sensitivity to nuclear shape}\label{sec:tri}

The configurations included in the configuration-mixing triaxial calculations for 
$^{45}$Sc/$^{45}$Cr mirror pair are depicted in Tab.~\ref{fig:Cr45PPT_CONF}. 
The table has conventional layout introduced in Ref.~\cite{(Lle20)}. Each configuration is represented by 
a set of symbols encoding occupation numbers.  Full dots denote pairwise occupied active Nilsson states while  up (down) arrows denote singly occupied active Nilsson states with signature $r=-i$ ($r=+i$), respectively.
The figure displays configurations in both members of the mirror pair, with left (right) column of symbols representing each configuration that  corresponds to the even (odd) subsystem, respectively. The configurations are grouped, 
also conventionally, into different types of excitation.  We adopt here the scheme worked out in Ref.~\cite{(Uth22)}
and divide them into the following groups: 
\begin{itemize}
    \item\textbf{g.s.} The first configuration listed in the table represents the global positive-parity ground state. 
    \item\textbf{ Group 1}: The simplest p-h seniority-one ($\nu =1$) excitations obtained by promoting the unpaired 
    nucleon to the empty active orbitals. 
    \item \textbf{Group 2}: Configurations that correspond to $\nu =1$ nn/pp pairing excitations
    \item \textbf{Group 3}: These are the lowest seniority-three 
($\nu =3$) configurations having one unpaired nucleon and one broken pair in the even subsystem.
    \item \textbf{Group 4}: These are the lowest seniority-three 
($\nu =3$) configurations having one unpaired nucleon and one broken pair in the odd subsystem.
    \item \textbf{Group 5}: These are seniority-three 
($\nu =3$) configurations having a hole in $sd$-shell in even subsystem.
\end{itemize}
For the positive parity $A$=45 mirrors there are 2 configurations of 
Group 1, 5 of Group 2, 8 of Group 3, 8 of Group 4 and 3 of Group 5. 
The overall number of configurations used here is comparable (only slightly larger) to the set 
used in Ref.~\cite{(Uth22)}. As before, each configuration, expressed in terms of occupation numbers in 
Tab.~\ref{fig:Cr45PPT_CONF}, is represented by a single self-consistent Slater determinant. The novel element 
is that we do not superimpose any constraint on the shape degree of freedom which, depending on convergence, 
can be either triaxial (T) or axial (A) as indicated in the table. 



\begin{table}[t]
\centering
\caption{(Color online)  
Configurations used in triaxial DFT-NCCI calculations for positive parity states in $^{45}$Sc/$^{45}$Cr
mirror nuclei. Left (right) column of symbols representing each configuration corresponds to the even (odd) subsystem, 
respectively i.e. the table displays mirror-symmetric sets of configurations in both nuclei. 
Full dots denote pairwise occupied Nilsson states.  Up (down) arrows denote singly occupied 
Nilsson states with signature $r=-i$ ($r=+i$), respectively.
Triaxial (axial) configurations are labeled as T (A), respectively.
}
\label{fig:Cr45PPT_CONF}
\begin{tabular}{|c|cc|cc|cc|cc|cc|cc|cc|cc|cc|cc|}\hline
GROUP & \multicolumn{2}{c|}{g.s.} & \multicolumn{4}{c|}{$1$} 
& \multicolumn{14}{c|}{3} \\\hline   
conf. no. & \multicolumn{2}{c|}{1} & \multicolumn{2}{c|}{2} & \multicolumn{2}{c|}{3} & \multicolumn{2}{c|}{17} & \multicolumn{2}{c|}{16} & \multicolumn{2}{c|}{4} & \multicolumn{2}{c|}{5} & \multicolumn{2}{c|}{19} & \multicolumn{2}{c|}{20} & \multicolumn{2}{c|}{21} \\\hline   
shape & \multicolumn{2}{c|}{T} & \multicolumn{2}{c|}{A} & \multicolumn{2}{c|}{A}
& \multicolumn{2}{c|}{T} & \multicolumn{2}{c|}{T} & \multicolumn{2}{c|}{T}
& \multicolumn{2}{c|}{T} & \multicolumn{2}{c|}{A} & \multicolumn{2}{c|}{A}
& \multicolumn{2}{c|}{A}  \\\hline   
$|303\: 7/2\rangle$ & 
& & 
& &
& &
& &
& &
& &
& &
& &
& &
& \\ 
$|312\: 5/2\rangle$ & 
& &
& &
& &
& &
& &
& &
& &
& &
& &
& \\ 
$|321\: 3/2\rangle$ & 
& \tikzcircle{0.7ex} &
& \tikzcircle{0.7ex} &
& \tikzcircle{0.7ex} &
{\color{blue} $\bm\uparrow$}     & \tikzcircle{0.7ex}                       &
{\color{blue} $\bm\downarrow$} & \tikzcircle{0.7ex}                       &
{\color{blue} $\bm\uparrow$}     & \tikzcircle{0.7ex}                       &
{\color{blue} $\bm\downarrow$} & \tikzcircle{0.7ex}                       &
{\color{blue} $\bm\uparrow$}     & \tikzcircle{0.7ex}                       &
{\color{blue} $\bm\downarrow$} & \tikzcircle{0.7ex}                       &
{\color{blue} $\bm\uparrow$}     & \tikzcircle{0.7ex}                       \\ 
$|330\: 1/2\rangle$ & 
\tikzcircle[blue, fill=blue]{0.7ex}  & \tikzcircle{0.7ex} &
\tikzcircle[blue, fill=blue]{0.7ex}  & \tikzcircle{0.7ex} &
\tikzcircle[blue, fill=blue]{0.7ex}  & \tikzcircle{0.7ex} &
{\color{blue} $\bm\uparrow$}     & \tikzcircle{0.7ex}                       &
{\color{blue} $\bm\uparrow$}     & \tikzcircle{0.7ex}                       &
{\color{blue} $\bm\downarrow$} & \tikzcircle{0.7ex}                       &
{\color{blue} $\bm\downarrow$} & \tikzcircle{0.7ex}                       &
{\color{blue} $\bm\uparrow$}     & \tikzcircle{0.7ex}                       &
{\color{blue} $\bm\uparrow$}     & \tikzcircle{0.7ex}                       &
{\color{blue} $\bm\downarrow$} & \tikzcircle{0.7ex}                       \\\hline 
$|202\: 3/2\rangle$ & 
{\color{blue} $\bm\uparrow$}      & \tikzcircle{0.7ex} &
\tikzcircle[blue, fill=blue]{0.7ex}   & \tikzcircle{0.7ex} &
\tikzcircle[blue, fill=blue]{0.7ex}   & \tikzcircle{0.7ex} &
{\color{blue} $\bm\uparrow$}      & \tikzcircle{0.7ex} &
{\color{blue} $\bm\uparrow$}      & \tikzcircle{0.7ex} &
{\color{blue} $\bm\uparrow$}      & \tikzcircle{0.7ex} &
{\color{blue} $\bm\uparrow$}      & \tikzcircle{0.7ex} &
\tikzcircle[blue, fill=blue]{0.7ex}   & \tikzcircle{0.7ex} &
\tikzcircle[blue, fill=blue]{0.7ex}   & \tikzcircle{0.7ex} &
\tikzcircle[blue, fill=blue]{0.7ex}   & \tikzcircle{0.7ex} \\  
$|200\: 1/2\rangle$ & 
\tikzcircle[blue, fill=blue]{0.7ex}   & \tikzcircle{0.7ex} &
{\color{blue} $\bm\downarrow$}  & \tikzcircle{0.7ex} &
\tikzcircle[blue, fill=blue]{0.7ex}   & \tikzcircle{0.7ex} &
\tikzcircle[blue, fill=blue]{0.7ex}   & \tikzcircle{0.7ex} &
\tikzcircle[blue, fill=blue]{0.7ex}   & \tikzcircle{0.7ex} &
\tikzcircle[blue, fill=blue]{0.7ex}   & \tikzcircle{0.7ex} &
\tikzcircle[blue, fill=blue]{0.7ex}   & \tikzcircle{0.7ex} &
{\color{blue} $\bm\downarrow$}  & \tikzcircle{0.7ex} &
{\color{blue} $\bm\downarrow$}  & \tikzcircle{0.7ex} &
{\color{blue} $\bm\downarrow$}  & \tikzcircle{0.7ex} \\  
$|211\: 1/2\rangle$ & 
\tikzcircle[blue, fill=blue]{0.7ex}   & \tikzcircle{0.7ex} &
\tikzcircle[blue, fill=blue]{0.7ex}   & \tikzcircle{0.7ex} &
{\color{blue} $\bm\downarrow$}  & \tikzcircle{0.7ex} &
\tikzcircle[blue, fill=blue]{0.7ex}   & \tikzcircle{0.7ex} &
\tikzcircle[blue, fill=blue]{0.7ex}   & \tikzcircle{0.7ex} &
\tikzcircle[blue, fill=blue]{0.7ex}   & \tikzcircle{0.7ex} &
\tikzcircle[blue, fill=blue]{0.7ex}   & \tikzcircle{0.7ex} &
\tikzcircle[blue, fill=blue]{0.7ex}   & \tikzcircle{0.7ex} &
\tikzcircle[blue, fill=blue]{0.7ex}   & \tikzcircle{0.7ex} &
\tikzcircle[blue, fill=blue]{0.7ex}   & \tikzcircle{0.7ex} \\\hline  


GROUP & \multicolumn{2}{c|}{3}  & \multicolumn{16}{c|}{4}  & \multicolumn{2}{c|}{5} \\\hline   
conf. no. & \multicolumn{2}{c|}{22} & \multicolumn{2}{c|}{7} & \multicolumn{2}{c|}{6} & \multicolumn{2}{c|}{13} & \multicolumn{2}{c|}{12} & \multicolumn{2}{c|}{23} & \multicolumn{2}{c|}{24} & \multicolumn{2}{c|}{25} & \multicolumn{2}{c|}{26} & \multicolumn{2}{c|}{18} \\\hline   
shape & \multicolumn{2}{c|}{} & \multicolumn{2}{c|}{T} & \multicolumn{2}{c|}{T}
& \multicolumn{2}{c|}{T} & \multicolumn{2}{c|}{T} & \multicolumn{2}{c|}{A}
& \multicolumn{2}{c|}{A} & \multicolumn{2}{c|}{A} & \multicolumn{2}{c|}{A}
& \multicolumn{2}{c|}{T}  \\\hline   
$|303\: 7/2\rangle$ & 
& & 
& &
& &
& &
& &
& &
& &
& &
& &
& \\ 
$|312\: 5/2\rangle$ & 
& &
& {\color{red} $\bm\downarrow$} &
& {\color{red} $\bm\uparrow$}     &
& {\color{red} $\bm\uparrow$}     &
& {\color{red} $\bm\downarrow$} &
& {\color{red} $\bm\downarrow$} &
& {\color{red} $\bm\uparrow$}     &
& {\color{red} $\bm\uparrow$}     &
& {\color{red} $\bm\downarrow$} &
& {\color{red} $\bm\uparrow$}    \\ 
$|321\: 3/2\rangle$ & 
{\color{blue} $\bm\downarrow$} & \tikzcircle{0.7ex} &
& {\color{red} $\bm\downarrow$} &
& {\color{red} $\bm\downarrow$} &
& {\color{red} $\bm\uparrow$}     &
& {\color{red} $\bm\uparrow$}     &
& {\color{red} $\bm\downarrow$} &
& {\color{red} $\bm\downarrow$} &
& {\color{red} $\bm\uparrow$}     &
& {\color{red} $\bm\uparrow$}     &
& \tikzcircle{0.7ex}                      \\ 
$|330\: 1/2\rangle$ & 
{\color{blue} $\bm\downarrow$} & \tikzcircle{0.7ex} &
\tikzcircle[blue, fill=blue]{0.7ex}  & \tikzcircle{0.7ex} &
\tikzcircle[blue, fill=blue]{0.7ex}  & \tikzcircle{0.7ex} &
\tikzcircle[blue, fill=blue]{0.7ex}  & \tikzcircle{0.7ex} &
\tikzcircle[blue, fill=blue]{0.7ex}  & \tikzcircle{0.7ex} &
\tikzcircle[blue, fill=blue]{0.7ex}  & \tikzcircle{0.7ex} &
\tikzcircle[blue, fill=blue]{0.7ex}  & \tikzcircle{0.7ex} &
\tikzcircle[blue, fill=blue]{0.7ex}  & \tikzcircle{0.7ex} &
\tikzcircle[blue, fill=blue]{0.7ex}  & \tikzcircle{0.7ex} &
{\color{blue} $\bm\downarrow$} & \tikzcircle{0.7ex} \\\hline 
$|202\: 3/2\rangle$ & 
\tikzcircle[blue, fill=blue]{0.7ex}   & \tikzcircle{0.7ex} &
{\color{blue} $\bm\uparrow$}      & \tikzcircle{0.7ex} &
{\color{blue} $\bm\uparrow$}      & \tikzcircle{0.7ex} &
{\color{blue} $\bm\uparrow$}      & \tikzcircle{0.7ex} &
{\color{blue} $\bm\uparrow$}      & \tikzcircle{0.7ex} &
\tikzcircle[blue, fill=blue]{0.7ex}   & \tikzcircle{0.7ex} &
\tikzcircle[blue, fill=blue]{0.7ex}   & \tikzcircle{0.7ex} &
\tikzcircle[blue, fill=blue]{0.7ex}   & \tikzcircle{0.7ex} &
\tikzcircle[blue, fill=blue]{0.7ex}   & \tikzcircle{0.7ex} &
\tikzcircle[blue, fill=blue]{0.7ex}   & {\color{red} $\bm\uparrow$} \\  
$|200\: 1/2\rangle$ & 
{\color{blue} $\bm\downarrow$}  & \tikzcircle{0.7ex} &
\tikzcircle[blue, fill=blue]{0.7ex}   & \tikzcircle{0.7ex} &
\tikzcircle[blue, fill=blue]{0.7ex}   & \tikzcircle{0.7ex} &
\tikzcircle[blue, fill=blue]{0.7ex}   & \tikzcircle{0.7ex} &
\tikzcircle[blue, fill=blue]{0.7ex}   & \tikzcircle{0.7ex} &
{\color{blue} $\bm\uparrow$}      & \tikzcircle{0.7ex} &
{\color{blue} $\bm\uparrow$}      & \tikzcircle{0.7ex} &
{\color{blue} $\bm\uparrow$}      & \tikzcircle{0.7ex} &
{\color{blue} $\bm\uparrow$}      & \tikzcircle{0.7ex} &
\tikzcircle[blue, fill=blue]{0.7ex}   & \tikzcircle{0.7ex} \\  
$|211\: 1/2\rangle$ & 
\tikzcircle[blue, fill=blue]{0.7ex}   & \tikzcircle{0.7ex} &
\tikzcircle[blue, fill=blue]{0.7ex}   & \tikzcircle{0.7ex} &
\tikzcircle[blue, fill=blue]{0.7ex}  & \tikzcircle{0.7ex} &
\tikzcircle[blue, fill=blue]{0.7ex}   & \tikzcircle{0.7ex} &
\tikzcircle[blue, fill=blue]{0.7ex}   & \tikzcircle{0.7ex} &
\tikzcircle[blue, fill=blue]{0.7ex}   & \tikzcircle{0.7ex} &
\tikzcircle[blue, fill=blue]{0.7ex}   & \tikzcircle{0.7ex} &
\tikzcircle[blue, fill=blue]{0.7ex}   & \tikzcircle{0.7ex} &
\tikzcircle[blue, fill=blue]{0.7ex}   & \tikzcircle{0.7ex} &
\tikzcircle[blue, fill=blue]{0.7ex}   & \tikzcircle{0.7ex} \\\hline  

GROUP & \multicolumn{4}{c|}{5} & \multicolumn{10}{c|}{2} & \multicolumn{6}{c|}{} \\\hline   
conf. no. & \multicolumn{2}{c|}{14} & \multicolumn{2}{c|}{15}  & \multicolumn{2}{c|}{9} & \multicolumn{2}{c|}{8} & \multicolumn{2}{c|}{10} & \multicolumn{2}{c|}{27} & \multicolumn{2}{c|}{28} & \multicolumn{2}{c|}{} & \multicolumn{2}{c|}{} & \multicolumn{2}{c|}{}  \\\hline  
shape & \multicolumn{2}{c|}{T} & \multicolumn{2}{c|}{T} & \multicolumn{2}{c|}{T} & \multicolumn{2}{c|}{T} & \multicolumn{2}{c|}{T} & \multicolumn{2}{c|}{A} & \multicolumn{2}{c|}{A} & \multicolumn{2}{c|}{} & \multicolumn{2}{c|}{} & \multicolumn{2}{c|}{} \\\hline   
$|303\: 7/2\rangle$ & 
& & 
& &
& &
& &
& &
& &
& &
& &
& &
& \\ 
$|312\: 5/2\rangle$ & 
& {\color{red} $\bm\uparrow$}      &
& {\color{red} $\bm\downarrow$}  &
& &
& \tikzcircle{0.7ex} &
\tikzcircle[blue, fill=blue]{0.7ex}  & &
& \tikzcircle{0.7ex} &
& &
& &
& &
& \\ 
$|321\: 3/2\rangle$ & 
                                                & \tikzcircle{0.7ex} &
                                                & \tikzcircle{0.7ex} &                                               
\tikzcircle[blue, fill=blue]{0.7ex}  & \tikzcircle{0.7ex} &
& &
                                                & \tikzcircle{0.7ex} &
& &
\tikzcircle[blue, fill=blue]{0.7ex}  & \tikzcircle{0.7ex} &
& &
& &
& \\ 
$|330\: 1/2\rangle$ & 
{\color{blue} $\bm\uparrow$}     & \tikzcircle{0.7ex} &
{\color{blue} $\bm\uparrow$}     & \tikzcircle{0.7ex} &
                                                & \tikzcircle{0.7ex} &
\tikzcircle[blue, fill=blue]{0.7ex}  & \tikzcircle{0.7ex} &
                                                & \tikzcircle{0.7ex} &
\tikzcircle[blue, fill=blue]{0.7ex}  & \tikzcircle{0.7ex} &
                                                & \tikzcircle{0.7ex} &
& &
& &
& \\\hline 
$|202\: 3/2\rangle$ & 
\tikzcircle[blue, fill=blue]{0.7ex}   & {\color{red} $\bm\uparrow$}      &
\tikzcircle[blue, fill=blue]{0.7ex}   & {\color{red} $\bm\uparrow$}      &
{\color{blue} $\bm\uparrow$}      & \tikzcircle{0.7ex} &
{\color{blue} $\bm\uparrow$}      & \tikzcircle{0.7ex} &
{\color{blue} $\bm\uparrow$}      & \tikzcircle{0.7ex} &
\tikzcircle[blue, fill=blue]{0.7ex}   & \tikzcircle{0.7ex} &
\tikzcircle[blue, fill=blue]{0.7ex}   & \tikzcircle{0.7ex} &
& &
& &
& \\  
$|200\: 1/2\rangle$ & 
\tikzcircle[blue, fill=blue]{0.7ex}   & \tikzcircle{0.7ex} &
\tikzcircle[blue, fill=blue]{0.7ex}   & \tikzcircle{0.7ex} &
\tikzcircle[blue, fill=blue]{0.7ex}   & \tikzcircle{0.7ex} &
\tikzcircle[blue, fill=blue]{0.7ex}   & \tikzcircle{0.7ex} &
\tikzcircle[blue, fill=blue]{0.7ex}   & \tikzcircle{0.7ex} &
{\color{blue} $\bm\downarrow$}  & \tikzcircle{0.7ex} &
{\color{blue} $\bm\downarrow$}  & \tikzcircle{0.7ex} &
& &
& &
& \\  
$|211\: 1/2\rangle$ & 
\tikzcircle[blue, fill=blue]{0.7ex}   & \tikzcircle{0.7ex} &
\tikzcircle[blue, fill=blue]{0.7ex}   & \tikzcircle{0.7ex} &
\tikzcircle[blue, fill=blue]{0.7ex}   & \tikzcircle{0.7ex} &
\tikzcircle[blue, fill=blue]{0.7ex}   & \tikzcircle{0.7ex} &
\tikzcircle[blue, fill=blue]{0.7ex}   & \tikzcircle{0.7ex} &
\tikzcircle[blue, fill=blue]{0.7ex}   & \tikzcircle{0.7ex} &
\tikzcircle[blue, fill=blue]{0.7ex}   & \tikzcircle{0.7ex} &
& &
& &
& \\\hline  

\end{tabular}
\end{table}

\begin{figure}[ht!]
\centering
\includegraphics[scale=0.50]{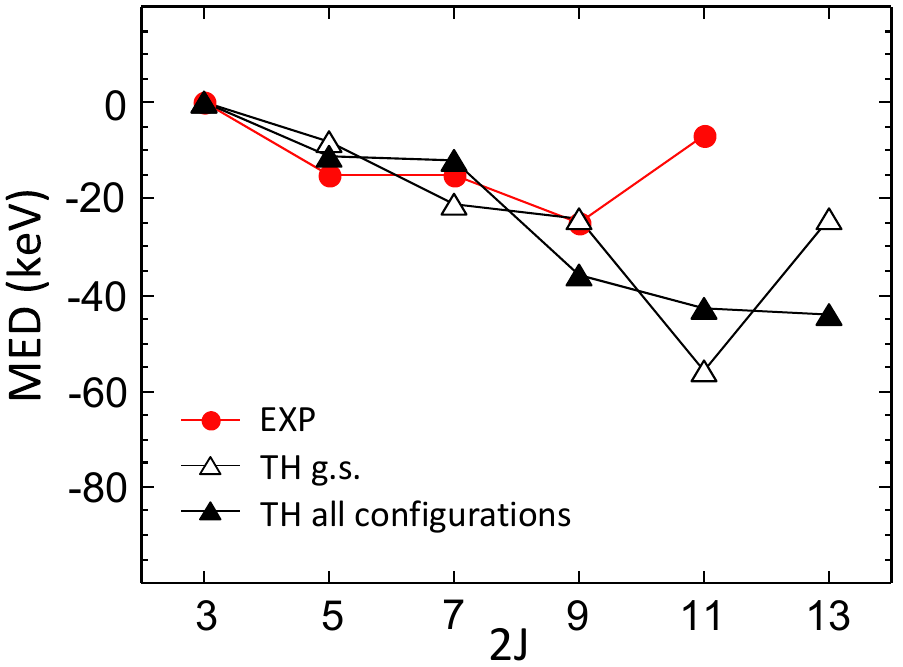}
\caption{(Color online) MED for the positive parity states in $^{45}$Sc/$^{45}$Cr in function of spin $2J$. 
Experimental data are marked with full (red) dots. Open triangles denote theoretical MED calculated using a single 
configuration representing g.s. Full (black) triangles represent the result of NCCI calculations involving all configurations
listed in Tab.~\ref{fig:Cr45PPT_CONF}.} 
\label{fig:Fig02}
\end{figure}

The unconstraint mean-field calculation leads to the g.s. configuration in $^{45}$Sc ($^{45}$Cr)
which is slightly triaxial with $\beta_2\approx 0.29 $ and $\gamma \approx 6.8^\circ$(6.3$^\circ$),
respectively. The triaxial minimum, however, is only 218\,keV  ($^{45}$Sc) and 134\,keV ($^{45}$Cr)
deeper as compared to the axially deformed g.s. configuration used in Ref.~\cite{(Uth22)}.
This indicates pronounced shape softness, at least in the low-spin part of the band. 
Inclusion of shape vibrations is prohibitively difficult with the present version of our model. Hence, 
as already discussed, we will represent the g.s. with a single triaxial self-consistent Slater determinant.

\begin{figure}[ht!]
\centering
\includegraphics[scale=0.50]{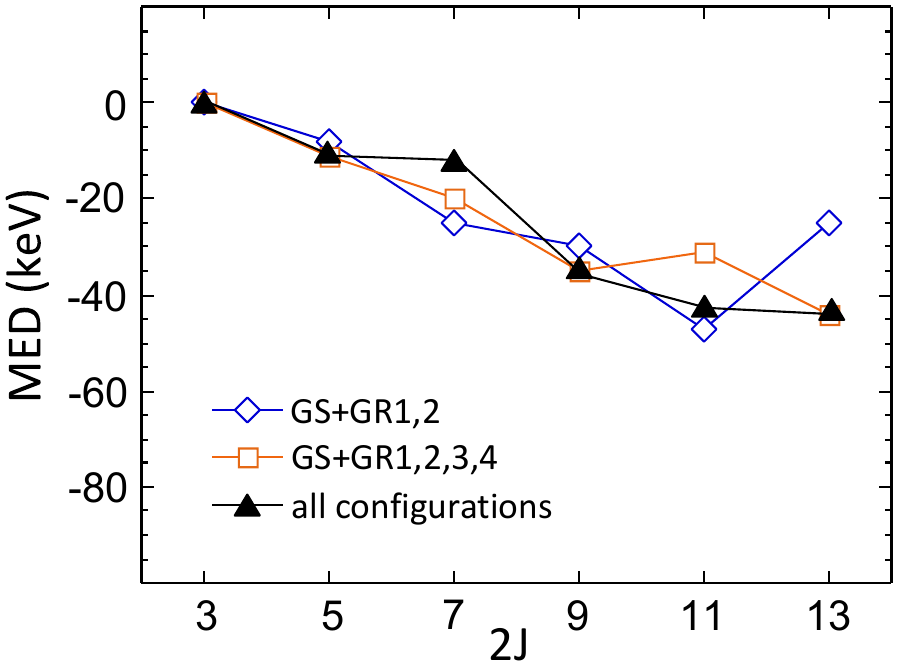}
\caption{(Color online) Theoretical MED for the positive parity states in $^{45}$Sc/$^{45}$Cr in function of spin $2J$. 
 Diamonds (blue) mark theoretical MED calculated using 
the g.s. and seniority $\nu =1$ excitations of groups 1 and 2. Squares (orange) include additionally the lowest 
$\nu =3$ excitations of groups 3 and 4. Full (black) triangles represent the result of NCCI calculations involving 
all configurations listed in Tab.~\ref{fig:Cr45PPT_CONF}.} 
\label{fig:Fig03}
\end{figure}

The calculation shows that triaxiality has a profound impact on the theoretical MED, in particular inverting their sign. 
Already by applying the angular-momentum projection to the g.s., without invoking configuration mixing, 
we obtain acceptably good agreement to data both concerning the sign as well as the magnitude of MED  
as shown in Fig.~\ref{fig:Fig02}. Interestingly this single-configuration-based picture rather weakly changes 
after admixing excited configurations what is rather atypical. This is visualized in Fig.~\ref{fig:Fig03} where
we show an impact of specific groups of excitations on MED. Indeed, the admixtures of the 
excited configurations of group 1 and 2 (marked by diamonds) almost do not change the 
MED obtained with the g.s. configuration alone shown (open triangles) in Fig.~\ref{fig:Fig02}.
The contributions due to the lowest seniority $\nu=3$ configurations of groups 3 and 4 tend to cancel 
each other. Their net effect on MED is small as shown (curve marked by squares)  in Fig.~\ref{fig:Fig03}.
The higher excitations of group 5  do not bring anything new (maybe except of spin 11/2$^+$) indicating
that our calculations are relatively well converged at low spins.   The overall agreement to the data 
should be considered as satisfactory although we are not able to account for sudden decrease 
at $J=11/2^+$ and subsequent increase at $J=13/2^+$ of the magnitude of MED.

\section{Summary and conclusions}\label{sec:sum}

Low-energy
nuclear processes including the superallowed $0^+ \rightarrow 0^+$~\cite{(Har20)} 
and $T=1/2$ mirror~\cite{(Nav09a),(Gon19)} beta decays allow for precision tests of fundamental symmetries.
The tests heavily rely on precision calculations of ISB corrections being a domain of many-body 
nuclear models. Hence, precision of nuclear tests of the Standard Model is heavily intertwined
with the credibility of nuclear modeling of ISB phenomena.   

During the last decade we have been developing a universal theoretical framework based on angular-momentum and 
isospin projected  DFT  to study ISB-related phenomena in $N\approx Z$ nuclei. The rationale behind choosing 
DFT-based method is their natural ability to account in a self-consistent way both the short- and long-range physics 
associated with strong and electromagnetic forces, respectively. The model appeared to be very successful in reproducing and 
predicting diverse observables and pseudo-observables associated with ISB effects from elusive isospin 
impurities~\cite{(Sat09)} to mirror energy differences (MED)~\cite{(Lle20),(Bac21)} in rotational bands 
versus angular momentum with the only exception of MED in the positive-parity mirror bands in 
$^{45}$Sc/$^{45}$Cr~\cite{(Uth22)}. In this case we failed to reproduce an overall sign of the MED
what casts a shadow on an overall good performance of our model and undermines its credibility.

In this work we performed a thorough analysis of the MED in $\pi = +$ bands of $^{45}$Sc/$^{45}$Cr 
mirror pair focusing on two aspects: 
({\it i\/}) sensitivity with respect to parameters of the effective contact CSB force 
and ({\it ii\/}) sensitivity with respect to nuclear shape. 
We have demonstrated that even large variations of the parameters of the contact 
CSB NLO force cannot invert sign of the calculated MED, at least not without deteriorating MDE for 
the $J=3/2^+$ band heads.  Our calculations reveal, on the other hand, 
strong dependence of the calculated MED on the shape of $\pi = +$ ground-state configuration.
The unconstraint Hartree-Fock solution for the $\pi=+$ g.s. configuration is triaxial and   
corresponds to $\beta_2 \approx 0.29$ and  $\gamma\approx 6.8^\circ$(6.3$^\circ$) in 
$^{45}$Sc ($^{45}$Cr), respectively. Inclusion of triaxial $\pi = +$ g.s. configuration
in the DFT calculations not only inverts the sign of  the calculated MED but also reproduces a 
magnitude of the MED without any need for further adjustment of the model's LECs. 
These two conclusions essentially do not depend on configuration mixing which weakly affects
the calculated MED. The physical mechanism staying behind such a radical change in dynamics 
of ISB effects along the rotational path is not clear and requires further studies.

\begin{acknowledgments}

This work was supported by the Polish National Science Centre (NCN) under Contract No 2018/31/B/ST2/02220. 

\end{acknowledgments}

\bibliographystyle{apsrev4-2}

\bibliography{biblio,MED,jacwit34}

\end{document}